\documentclass[sigconf]{acmart}
\usepackage{booktabs} 
\usepackage{multirow}
\usepackage{subcaption}
\usepackage{caption}
\usepackage{setspace}
\usepackage{amsmath}
\usepackage[english]{babel}
\definecolor{mygray}{gray}{0.5}
\usepackage{float}
\usepackage{flushend}
\usepackage{mathtools}

\definecolor{midnightgreen}{rgb}{0.0, 0.29, 0.33}
\definecolor{orange}{RGB}{255,127,0}

\newcommand{\ie}{\emph{i.e.}}
\newcommand{\eg}{\emph{e.g.}}
\newcommand{\wrt}{\emph{w.r.t. }}
\newcommand{\abbrv}{\emph{abbrv. }}

\newcommand*{\uset}[3][0pt]{%
  \begingroup
    \renewcommand*{\arraystretch}{0}%
    \:\begin{array}[t]{@{}c@{}}%
      #2\\[{#1}]%
      \scriptstyle #3%
    \end{array}\:%
  \endgroup
}

\copyrightyear{2019} 
\setcopyright{acmcopyright} 
\acmConference[SIGIR '19]{SIGIR 2019}{July 21-25, 2019} {Paris, France}

\begin{document}
\title{An Axiomatic Approach to Regularizing Neural Ranking Models}

\author{
  Corby Rosset, Bhaskar Mitra, Chenyan Xiong, Nick Craswell, Xia Song, and Saurabh Tiwary \\
  Microsoft AI \& Research \\
  \texttt{\{corosset, bmitra, cxiong, nickcr, xiaso, satiwary\}@microsoft.com}
}


\begin{abstract}
Axiomatic information retrieval (IR) seeks a set of principle properties desirable in IR models.
These properties when formally expressed provide guidance in the search for better relevance estimation functions.
Neural ranking models typically contain a large number of parameters.
The training of these models involve a search for appropriate parameter values based on large quantities of labeled examples.
Intuitively, axioms that can guide the search for better traditional IR models should also help in better parameter estimation for machine learning based rankers.
This work explores the use of IR axioms to augment the direct supervision from labeled data for training neural ranking models.
We modify the documents in our dataset along the lines of well-known axioms during training and add a regularization loss based on the agreement between the ranking model and the axioms on which version of the document---the original or the perturbed---should be preferred.
Our experiments show that the neural ranking model achieves faster convergence and better generalization with axiomatic regularization.

\end{abstract}
\maketitle
\section{Introduction}
\label{sec:intro}

The goal of axiomatic information retrieval (IR) \citep{fang2004formal, fang2005exploration, fang2011diagnostic} is to formalize a set of desirable constraints that any reasonable IR models should (at least partially) satisfy.
For example, one of the axioms (TFC1) states that a document containing more occurrences of a query term should receive a higher score.
According to another axiom (LNC1), extra occurrences of non-relevant terms should negatively impact the score of a document.
All else being equal, an IR model that satisfies these two axioms should theoretically be more effective than one that does not.
The formalization of these axioms, therefore, provide a means to analyse IR models analytically, in lieu of purely empirical comparisons.
As a corollary, these axioms can help in the search for better retrieval functions given a candidate space of IR models \citep{fang2005exploration}.

Most neural approaches to IR \citep{mitra2018introduction} consider models with large number of parameters.
The training procedure for these models typically involve an iterative search---\eg, using stochastic gradient descent \citep{bottou2010large}---to find good combinations of model parameters by leveraging large quantities of labeled data.
Intuitively, IR axioms---that can guide the search for models in the space of traditional IR methods---should also be useful in optimizing the parameters of neural IR models.
Under supervised settings, neural ranking models learn by comparing two (or more) documents for a given query and optimizing its parameters such that the more relevant document receives a higher score.
An over-parameterized model may find several ways to fit the training data.
But in the presence of many possible solutions, we hypothesize that it is preferable to find the solution that conforms to well known axioms of IR.

In this work we propose to incorporate IR axioms to regularize the training of neural ranking models.
We select five axioms---TFC1, TFC2, TFC3, TDC, and LNC---for this study, that we describe in more details in Section \ref{sec:model}.
We perturb the documents in our training data along the lines of these axioms.
For example, to perturb a document using TFC1 we add more instances of the query terms to the document.
During training---in addition to comparing documents of different relevance grades for a query---we also compare the documents to their perturbed version.
We compute a regularization loss based on the agreement (or disagreement) between the ranking model and the axiom on which version of the document---the original or the perturbed---should be preferred.

Our experiments show that axiomatic regularization is effective at speeding up convergence of neural IR models during training and achieves significant improvements in effectiveness metrics on heldout test data.
In particular, axiomatic regularization helps a simple yet effective neural learning to rank model, Conv-KRNM (CKNRM)~\cite{dai2018convolutional}, improve MRR on MS-MARCO and a large internal dataset by about 3\%.
The improvements from axiomatic regularization are particularly encouraging under the smaller training data regime---which indicates it may be useful in alleviating our dependence on the availability of large training corpus in neural IR.

\section{Related Work}
\label{sec:related}

\paragraph{Axiomatic IR}

While inductive analysis of IR models have been previously attempted \citep{bruza1994investigating}, it was \citet{fang2004formal} who proposed the original six IR axioms related to term frequency (TFC1 and TFC2), term discrimination (TDC), and document length normalization (LNC1, LNC2, and TF-LNC)---followed by an additional term frequency constraint (TFC3) by \citet{fang2011diagnostic}.
Since then these axioms have been further expanded to cover term proximity \citep{tao2007exploration}, semantic matching \citep{fang2006semantic, fang2008re}, and other retrieval aspects \citep{lv2011lower, zheng2010query, wu2012relation}.
We refer the reader to \citep{zhai2013axiomatic} for a more thorough review of the existing axioms.
Recently, \citet{renningsaxiomatic} adopted these axioms to analyze different neural ranking models.
However, this is the first study that leverages IR axioms to regularize neural ranker training.

\paragraph{Incorporating domain knowledge in supervised training}
State-of-the-art neural ranking models---\eg, \citep{dai2018convolutional, nogueira2019passage, mitra2017learning}---have tens of millions to hundreds of millions of parameters.
Models with such large parameter sets can overfit when only small amount of training data is available.
Domain knowledge may help identify additional sources of supervision, or inform methods for regularization to compensate for the lack of enough training data.
Weak supervision using domain knowledge has been effective in many application areas with little or no training data---including, entity extraction \citep{mintz2009distant}, computer vision \citep{stewart2017label}, and IR \citep{dehghani2017neural}.
In a supervised setting, data augmentation methods may be developed based on domain knowledge.
In computer vision a labeled image can be scaled, flipped or otherwise transformed in ways that create a different image, but the label is still valid \citep{perez2017effectiveness}. 
Similarly in machine translation, data can be augmented by replacing words on both sides of a training pair, while tending to preserve a valid translation \citep{fadaee2017data}.  
A different approach is to incorporate domain knowledge as a regularizer. For example, when predicting a physical response, adding a penalty term for diverging from laws of physics \citep{nabian2018physics}. In this study we adopt the regularization approach.





\section{Axiomatic regularization for neural ranking models}
\label{sec:model}

In ad-hoc retrieval---an important IR task---the ranking model receives as input a pair of query $q$ and document $d$, and estimates a score proportional to their mutual relevance.
The learning-to-rank literature \citep{Liu:2009} explores a number of loss functions that can be employed to discriminatively train such a ranking model $s_\theta$.
We use the hinge loss \citep{herbrich2000large} in this study.

\begin{align}
    \mathcal{L} &= \mathbb{E}_{q \sim \phi,\; d_{pos}, d_{neg} \sim \psi} [\ell(q, d_{pos}, d_{neg})]
    \label{eqn:loss-original} \\
    \ell(q, d_{pos}, d_{neg}) &= \text{max}\{0, \epsilon - \big(s_\theta(q, d_{pos}) - s_\theta(q, d_{neg})\big)\}
\end{align}

Minimizing the hinge loss implies maximizing the gap between $s_\theta(q, d_{pos})$ and $s_\theta(q, d_{neg})$---where query $q$ is sampled randomly from distribution $\phi$ and documents $d_{pos}$ and $d_{neg}$ from $\psi$.
We use the notation $d_{pos} \uset{\succ}{q} d_{neg}$ to denote that the document $d_{pos}$ is more relevant of the two documents \wrt query $q$.

We define a set $\Delta$ of axiomatic regularization constraints based on existing IR axioms.
Each regularization constraint $\Delta_i$ defines a dimension in which a document $d$ can be perturbed---to generate a new document $\mathrm{d}^{(i)}$---such that its relevance to a query $q$ is impacted---either positively or negatively.
Let $\delta_i \in \{+1, -1\}$ be equal to $1$, if the constraint $\Delta_i$ states that $d \uset{\succ}{q} \mathrm{d}^{(i)}$---\ie, the original document $d$ should be considered as more relevant than $\mathrm{d}^{(i)}$ \wrt query $q$---and be equal to $-1$ otherwise.

We redefine the hinge loss of Equation \ref{eqn:loss-original} to include the axiomatic regularization (\abbrv `AR') below.

\begin{align}
    \mathcal{L} &= \mathbb{E}_{q \sim \phi,\; d_{pos}, d_{neg} \sim \psi, \Delta_i \sim \upsilon} [\ell(q, d_{pos}, d_{neg})]
\end{align}

\begin{equation}
\begin{aligned}
    \ell_{AR}(q, &d_{pos}, d_{neg}, \Delta_i) =\\
    &\text{max}\{0, \epsilon - \big(s_\theta(q, d_{pos}) - s_\theta(q, d_{neg})\big)\} \\
    &+ \lambda \cdot \text{max}\{0, \mu - \delta_i \cdot \big(s_\theta(q, d_{pos}) - s_\theta(q, \mathrm{d}^{(i)}_{pos})\big)\} \\
    &+ \lambda \cdot \text{max}\{0, \mu - \delta_i \cdot \big(s_\theta(q, d_{neg}) - s_\theta(q, \mathrm{d}^{(i)}_{neg})\big)\}
    \label{eqn:loss-new}
\end{aligned}
\end{equation}

where, $\upsilon$ is the uniform distribution over all axiomatic regularization constraints in $\Delta$.
We treat $\lambda$ and $\mu$ as hyper-parameters.

In this study, we consider three of the standard IR axioms that we formally state below.
\begin{itemize}
    \item [TFC1] This axiom states that we should give higher score to a document that has more occurrences of a query term. \\
    if: $|q| = 1$, $|d_i| = |d_j|$, and $\#(q_1, d_i) > \#(q_1, d_j)$,\\
    then: $d_i \uset{\succ}{q} d_j$\\
    where, $\#(t, u)$ denotes the term frequency of $t$ in text $u$.
    \item [TFC3] This axiom states that if the cumulative term frequency of all query terms in both documents are same and every term is equally discriminative, then a higher score should be given to the document covering more unique terms.\\
    if: $|q| = 2$, $|d_i| = |d_j|$, td$(q_1)$ = td$(q_2)$, $\#(q_1, d_i) = \#(q_1, d_j) + \#(q_2, d_j)$, $\#(q_2, d_i) = 0$, $\#(q_1, d_j) \neq 0$, and $\#(q_2, d_j) \neq 0$,\\
    then: $d_j \uset{\succ}{q} d_i$\\
    where, td$(t)$ is any measure of term discrimination, such as inverse document frequency \citep{robertson2004understanding}.
    \item [LNC] This axiom states the score of a document should decrease if more non-relevant terms are added.\\
    if: $t \notin q$, $\#(t, d_j) = \#(t, d_i) + 1$, $\forall w \in d_i \cup d_j, \#(w, d_i) = \#(w, d_j)$,\\
    then: $d_j \uset{\not\succ}{q} d_i$\\
\end{itemize}

Based on these stated axioms we derive the set $\Delta$ of four regularization constraints.

\begin{itemize}
    \item [TFC1-A] We randomly sample a query term and insert it at a random positions in document $d$. We expect the perturbed document $\mathrm{d}^{(i)}$ to be more relevant to the query---\ie, $\mathrm{d}^{(i)} \uset{\succ}{q} d$.
    \item [TFC1-D] We randomly sample one query term and delete all its occurrences in document $d$. We expect the perturbed document to be less relevant to the query---\ie, $d \uset{\succ}{q} \mathrm{d}^{(i)}$.
    \item [TFC3] We randomly sample one of the query terms not present in document $d$, if any, and insert it at a random position in the document. We expect the perturbed document to be more relevant to the query---\ie, $\mathrm{d}^{(i)} \uset{\succ}{q} d$.
    \item [LNC] We randomly sample $k$ terms from the vocabulary and insert them at random positions in the document $d$. We expect the perturbed document to be less relevant to the query---\ie, $d \uset{\succ}{q} \mathrm{d}^{(i)}$.
\end{itemize}

Next, we describe our experiment methodology and present results from the empirical study.

\section{Experiments}
\label{sec:experiment}

For reproducibility, we use an open-source repository of neural ranking models\footnote{https://github.com/thunlp/Kernel-Based-Neural-Ranking-Models} containing CKNRM~\cite{dai2018convolutional}, which we train on the publicly available MS MARCO \citep{bajaj2016ms} ranking dataset\footnote{http://www.msmarco.org/}.
The train and dev set in MS MARCO contains 398,792 and 6,980 queries, respectively.
For each query, the top 1000 passages are retrieved by BM25.
On average, about one passage is manually labeled as relevant to the query. 

For the MS MARCO experiments we use the CKNRM model.
We use the 400K GloVe vocabulary\footnote{https://nlp.stanford.edu/projects/glove/} to initialize the word embeddings. 
The out-of-vocabulary rate was about 1\% on MS MARCO training and dev data. 

For training CKNRM, we use its default hyperparameters in the repository: learning rate 0.001, batch size 64, and Adam optimizer with weight decay.
We sub-sample 512 out of the 6,900 queries from the MS MARCO dev set to select the best model in intermediate evaluations during training, and then evaluate on the remaining dev queries.
We generate one perturbation of each of the positive and negative passages in each row of the MS MARCO training data by independently and uniformly at random choosing an axiom from \{TFC1-A, TFC1-D, TFC3, LNC\}.

We add to the original CKNRM ranking loss two additional axiomatic hinge losses: one comparing the pair of original and perturbed positive passage, and similarly for the pair of negative passages.
We tune the coefficient of the axiomatic loss, $\lambda$, and its margin, $\mu$, over \{0.001, 0.01, 0.1, 0.25, 0.5, 1.0\} and find that smaller coefficients and smaller margins work better as the size of the training dataset increases. 

To show how axiomatic regularization impacts learning, we train CKNRM and its axiomatic variant on four subsamples of the MS MARCO ranking dataset.
We sub-sample 100, 1k, 10k, and 100k queries from the data and include all the passage pairs for the subsampled queries.
We then train four independent models of the baseline CKNRM and its axiom-regularized variant on each of the datasets, and ensemble the models by averaging their scores for each document in the dev set to produce the MRR numbers shown in Figure \ref{fig:diff_dataset_sizses}.
Every model is trained for exactly 15,000 steps, except for the points on the far right, which are trained for 60,000 steps on all 300k queries of the MS MARCO training data. 

We perform an ablation study of adding each axiom in isolation to the original hinge loss of CKNRM in Table \ref{table:ablation}.

We also apply axiomatic regularization to a proprietary ranking dataset from a commercial search engine---comprised of 10 documents for each of 500k queries.
The documents have human judgments on the \{bad, fair, good, excellent, perfect\} scale.
There are two evaluation sets, a sample of about 16K queries from a six month period weighted by their occurrence in the log, and an unweighted (uniform) sample of queries from the same six month period.
We use a proprietary deep neural model (DNN) to encode the query and the document from its various fields including Title, URL, Anchor, and Co-Clicks.
The model is trained to regress to the pointwise relevance label using mean square error loss, to which we add the axiomatic regularization.
We compare this DNN model and its axiom-regularized variant to BM25 in Table \ref{table:msmarco_and_internal} (Bottom).

\section{Results}
\label{sec:results}

\begin{figure}
    \centering
    \includegraphics[scale=0.67]{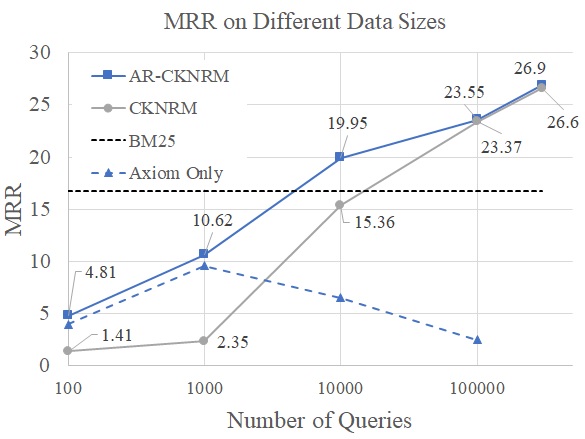}
    \caption{MRR results of training CKNRM and its axiomatic variant on datasets with 100, 1k, 10k, 100k, and all MS-MARCO queries on the dev set. Each point represents the ensemble of four independently trained models.}
    \label{fig:diff_dataset_sizses}
    \vspace{-4mm}
\end{figure}

We show the value of axiomatic regularization in Figure \ref{fig:diff_dataset_sizses} across a variety of data sizes subsampled from MS-MARCO. Its impact is most pronounced in low-data scenarios where it significantly improves a deep neural model that was struggling to capture basic relevance signals on 100, 1k, or even 10k query datasets. Only after introducing axiomatic regularization could CKNRM overtake BM25 on 10k queries. In fact, for these low volume datasets the best hyperparameters for the axiomatic loss were at least 0.25 for both the loss coefficient and the margin, suggesting that the axioms played a major role in guiding the model. 

\begin{figure}
    \centering
    \includegraphics[scale=0.63]{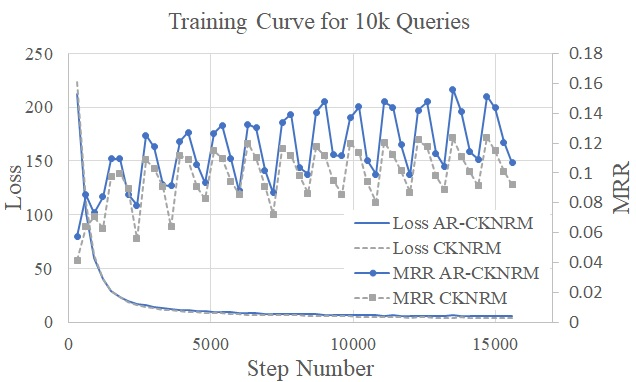}
    \caption{Training curve of the loss and dev MRR of both CKNRM and AR-CKNRM on the 10k query dataset}
    \label{fig:train_curve}
    \vspace{-4mm}
\end{figure}

These axiomatic hyperparameters transitioned lower, however, in the higher data scenarios which are more accommodating for neural models. This agrees with our intuition that regularization coefficients should contribute only a fraction of the total loss, and the margin separating a document and its perturbation should be smaller than that separating documents of different human-labeled relevance classes. The best empirical axiomatic hyperparameters agree with these intuitions; the coefficient and margins were all at or below 0.1. In this case, the axioms behaved more like traditional regularization techniques. We show the regularizing effect in Figure \ref{fig:train_curve}, where we plot the original hinge loss (without axiomatic loss added in) and the dev MRR for both types of models. 

\begin{table}[]
\begin{tabular}{rcc}
\cline{2-3}
\multicolumn{1}{c}{} & \multicolumn{2}{c}{\textbf{Results on MS-MARCO}}                  \\
\multicolumn{1}{l}{} & MAP                              & MRR                            \\
CKNRM                & 25.75                            & 26.07                          \\
AR-CKNRM             & 26.62                            & 26.94                          \\
\multicolumn{1}{l}{} & \multicolumn{1}{l}{}             & \multicolumn{1}{l}{}           \\ \cline{2-3} 
\multicolumn{1}{c}{} & \multicolumn{2}{c}{\textbf{Results (NDCG@1) on Proprietary Data}} \\
\multicolumn{1}{l}{} & Weighted                         & Unweighted                     \\
BM25                 & 33.69                            & 23.75                          \\
DNN                  & 44.04                            & 25.11                          \\
AR - DNN             & 45.39                            & 26.13                         
\end{tabular}
\caption{(Top) Results on the MS-MARCO Eval set of the ensemble of four models trained on all MS-MARCO queries. (Bottom) NDCG@1 numbers of a proprietary neural model and its axiomatic variant on an large scale commercial ranking dataset. (All values are x100)}
\label{table:msmarco_and_internal}
\vspace{-4mm}
\end{table}

Even when data is abundant, where deep models typically thrive, Table \ref{table:msmarco_and_internal} (Top) demonstrates that the axioms still contribute noticeable improvements which are competitive with the MS MARCO leaderboard. On the MS MARCO eval dataset, axiomatic regularization improves performance by about 3\%. 
This improvement is also consistent with that of NDCG on the proprietary ranking dataset in Table \ref{table:msmarco_and_internal} (Bottom). 

\begin{table}[]
\begin{tabular}{rcc}
\multicolumn{3}{c}{Ablation on 10k Queries}      \\ \hline
                                 & MAP   & MRR   \\
CKNRM                            & 15.13 & 15.36 \\
+ TFC1-A                         & 19.33 & 19.56 \\
+ TFC1-D                         & 18.16 & 18.38 \\
+ TFC3                           & 19.05 & 19.28 \\
+ LNC                           & 11.42 & 11.47 \\
\multicolumn{1}{l}{+ All Axioms} & 19.70 & 19.95
\end{tabular}
\caption{An add-one-in ablation study of each of the axiomatic losses; the last row shows all axioms.}
\label{table:ablation}
\vspace{-8mm}
\end{table}
Table \ref{table:ablation} shows the results of an add-one-in ablation study of each axiom added individually to the original hinge loss. On their own, TFC1 and TFC3 are enough to provide a roughly 30\% relative improvement on a dataset of 10k queries, reinforcing the importance of query term matching signals which CKNRM on its own could not capture. Curiously, however, LNC1 on its own hinders performance, which raises the question of how to best teach a neural model to penalize noise terms and length of a document.

\section{Conclusion}
\label{sec:conclusion}

While some traditional IR methods have directly inspired specific neural architectures---\eg, \citep{zamani2018neural}---arguably much of neural IR's current recipes have been borrowed from other application areas of deep learning, such as natural language processing.
It is therefore exciting to see a framework like axiomatic IR---that was originally intended to provide an analytical foundation for classical retrieval methods---proving effective in improving generalizability of modern neural approaches.
While we find axiomatic constraints to be effective as regularization schemes, we suspect they may also hold the key to thinking about novel unsupervised and distant learning strategies for IR tasks.


\bibliographystyle{ACM-Reference-Format}
\bibliography{citation}
\end{document}